\documentclass{elsarticle}
 \usepackage{lineno,hyperref}
 \modulolinenumbers[5]

\journal{Journal of New Astronomy}

\begin{document}

\begin{frontmatter}

\title{General model of  depolarization and transfer of polarization  of singly ionized atoms by collisions with hydrogen atoms}

 \author{M. Derouich$^{1,2}$}
\address{$^{1}$Astronomy Department, Faculty of Science, King Abdulaziz University, 21589 Jeddah, Saudi Arabia\\
$^{2}$Sousse University,  ESSTHS, Lamine Abbassi street, 4011 H. Sousse, Tunisia}


\cortext[mycorrespondingauthor]{M. Derouich}
\ead{derouichmoncef@gmail.com}


\begin{abstract}
 Simulations of the generation of the atomic polarization is necessary for interpreting the  second solar spectrum.  For this purpose, it is important to rigorously determine   the  effects of the isotropic collisions  with neutral hydrogen on the atomic polarization of the neutral atoms, ionized atoms and molecules.   
 Our aim is to treat    in generality the problem of   depolarizing isotropic collisions between singly ionized atoms and neutral hydrogen in its ground state.  
Using our numerical code, we computed the collisional depolarization rates of  the $p$-levels of ions for large number of values of the effective principal quantum number $n^{*}$ and the Uns\"old energy $E_p$. Then, genetic programming  has been utilized to fit the  available    depolarization rates. As a result, strongly non-linear relationships between the collisional depolarization rates,   $n^{*}$ and  $E_p$ are obtained, and are shown to reproduce the original data with accuracy clearly better than 10\%. These relationships   allow  quick calculations of the depolarizing collisional rates of any simple ion which is very useful for the solar physics community. In addition, the depolarization rates associated to the complex ions and to the hyperfine  levels can be easily derived from our results.     
 In this work we have shown that by using powerful  numerical approach and our   collisional method, general model giving the depolarization of the ions can be obtained to be exploited  for solar applications.  
\end{abstract}


\end{frontmatter}


\section{Introduction}
The transitions between levels of solar ions  caused by    anisotropic scattering of the incident   radiation
 induce population imbalances and quantum coherences among the Zeeman sublevels. 
 Population imbalances and quantum coherences are usually called  {\it atomic polarization}.
  The  second solar spectrum  (SSS), which is the spectrum of the linear polarization observed near the limb of the Sun, is the observational signature of the atomic polarization.  The term SSS  was coined by Ivanov (1991). 
 It is also usually referred to as spectrum of the scattering polarization since it is due to coherent scattering processes (e.g. Trujillo Bueno  2001).  
 
 The SSS is modified by solar magnetic fields via the Hanle effect. Such a modification    allows   diagnostics of the magnetic fields.  In fact, Hanle effect   diagnostics  of hidden, mixed-polarity magnetic fields at sub-telescopic scales require confronting  the discrepancy between the polarization calculated in the absence of    magnetic fields and the  observed  linear scattering polarization (e. g. Stenflo 1982;  Landi Degl'Innocenti  1983; Sahal-Br\'echot et al 1986; Stenflo 2004; Trujillo Bueno et al. 2004; Derouich et al. 2006; Faurobert et al. 2009). Reliable diagnostics  consists in solving     numerically   the coupled set of  equations of the radiative transfer and the statistical equilibrium   of a multilevel  atomic system  taking into account radiative and collisional processes.   The collisional rates, which are often   poorly known,  are a fundamental ingredient for   realistic diagnostics.  Sometimes the information encoded in accurate solar sspectropolarimetric   observations would be inaccessible if the effect of the collisions is unknown (e.g.  Derouich et al. 2006; Derouich et al. 2007).  There is a need for theory and numerical modeling of collisional depolarization of spectral lines by  nearby hydrogen atoms. We notice that the collisions with  hydrogen are dominant because it is the most abundant atom in the solar photosphere where the SSS is formed.

 In this context, we aim to generalize our numerical modeling of Derouich et al. (2004) to provide collisional data for any  ionic   level. 
The paper is organized as follows. Section 2 gives a brief review of the theoretical background. In Section 3, we formulate the problem.  Section 4 explains the basic definitions and notations,   describes the collisional data employed in the context of this work, and provides the results for simple atoms without hyperfine structure. Simple atoms with hyperfine structure and complex atoms  are presented in sections   5 and 6. Finally, the conclusion of the paper is presented in section 7.

\section{Brief theoretical background}
In real plasmas like the solar atmosphere, emitting ions suffer the effects of  collisions with nearby abundant particles.    To correctly extract informations contained in the solar observations would be inaccessible,  the effect of the collisions must be taken into account.  During the 2000s, Derouich, Sahal-Br\'echot and Barklem (DSB)  developed a semi-classical theory allowing   precise and quick calculations of    the depolarization and polarization transfer rates by collisions with neutral hydrogen  (see  for example    Derouich et al. (2003a,b,   2005a,b); Derouich 2004; Derouich and Barklem (2007)). The   DSB approach is based on the  Anstee-Barklem-O'Mara (ABO) theory concerned with   collisional line broadening by neutral hydrogen  (Anstee 1992; Anstee \& O'Mara 1991, 1995; Anstee et al. 1997; Barklem 1998; Barklem \& O'Mara 1997; Barklem et al. 1998). 

An important result which justifies
the use  of the DSB semi-classical approach is that the collisional depolarization and polarization transfer probabilities depend 
exclusively on the intermediate range of the interatomic separations. We have shown that  
a modification of the interaction potential values by a Gaussian magnification factor,  for the close  or   
long-range regions of the interaction, does not practically change the values of 
the calculated collisional probabilities (Derouich et al. 2003a, 2004).  In fact, we have found that a significant effect on the  collisional depolarization and polarization transfer probabilities takes place  only in the intermediate range of the interatomic separations. The sophisticated quantum chemistry  approach is   different to our  semi-classical approach only in the short-range regions of the interactions. Since  close interaction regions do not influence the values of the depolarization rates, we
 obtain  good results compared to the available quantum chemistry ones ($\sim$ 10 \% of accuracy). 
 
 In addition,  the semi-classical approach is very useful for   heavy and/or complex atoms/ions    like Fe I, Ti I,  NdII, EuII, CeII, ZrII, etc.., whose collisional rates cannot be presently obtained via   quantum chemistry methods. The spectral lines of such atoms/ions   show significant polarization peaks in many spectral lines (see the atlases by Gandorfer 2000, 2002, 2005).   
 
 Although the fact that this theory is of semi-classical nature, the close coupling is taken into account and the  time-dependent Schr\"odinger equations 
is solved. 
 
\section{Formulation of the problem}
For neutral atoms,  the DSB and ABO theories give general results for  neutral atoms however they lose their generality in the case of ionized atoms. This is beacause, for all neutral atoms,  the Uns\"old energy $E_p$ can be replaced   by a constant value 
 E$_p$ = -4/9 atomic units (see, e.g., Anstee 1992, Barklem \& O'Mara 1998; Derouich et al.  2003a).     
 
On the contrary,  while ABO and DSB theories can be applied for any singly ionized atom, the 
 calculation of the parameter  E$_p$ is required for each level which implies that the results are specific for each level of a particular   ion.
The interaction energy associated to the interaction of a singly ionized atom  
with a hydrogen atom in its ground state is given by Equation (4) of Derouich et al. (2004).
After some derivations and  by using the Uns\"old approximation, the expression of the interaction potential 
becomes dependent on the paramater E$_p$ as shown by Equation (9) of Derouich et al. (2004). 
The variation of the depolarization rates as a function of  E$_p$ is given by Derouich et al. (2004) who concluded that 
one has to determine E$_p$ before going to the calculation  of the depolarization rates.  
This limitation   substantially restrict the generalization of the  results of Derouich et al. (2004).
In fact,  two steps are needed in the calculation of the depolarization rates for the levels of ions
\begin{enumerate}
\item Step 1:  
one must determine Ep directly for each state of 
ech  ion  via the expression: 
\begin{eqnarray} \label{eq_1}
E_p = - \frac{2<p^2_2>} {C_6}, 
\end{eqnarray}
 where $<p^2_2>$ is the mean square distance between the optical electron and the perturbed ion core, 
 \begin{eqnarray} \label{eq25} 
<p^2_2>= \frac{n^{*2}}{2Z^2} [5n^{*2}+1-3l(l+1)].
\end{eqnarray}
$l$=1 is the orbital momentum for $p$-states, $Z$=2 for singly ionized atoms and $n^{*}$ is the effective quantum number (see Derouich et al. 2004). $C_6$ is the Van der Waals coefficient  given by the standard relationship (see for intstance  Goodisman 1973 and Derouich et al. 2004) 
\item Step 2:  Then, the value of E$_p$ is included in the expression of the interaction energy.  After that, the probabilities of collisions are obtained by solving  the semi-classical differential coupled equations which are derived from the time-dependent Schr\"odinger equation (Derouich et al. 2003a). To obtain the depolarization rates, one thus must perform the  integration of the probabilities of collisions 
over the impact-parameter $b$ and over a Maxwell distribution 
of velocities $f(v)$ for a temperature $T$ of the medium. 
\end{enumerate}
In practice,   solar physicists can calculate the value of $E_p$ (step 1) but it might be quite complicated  for them to determine the depolarization rates  by using the  collision theory (step 2). The main goal of this  work is 
to  overcome numerically the difficulty pointed out  in the step 2 to give the possibility of the determination of the depolarization rates by completing only step 1.

\section{Definitions, notations and numerical results}
\subsection{Definitions and notations}

We denote the   atomic levels by ($\alpha$ $J$)     where   $J$ is the total angular momentum  of the level and  $\alpha$ represents 
 the other quantum numbers necessary to  define the electronic level of the ion.    The atomic states ($\alpha$ $J$) are described  by the   tensorial components $\rho_{q}^{k} (\alpha J)$  of the atomic density matrix.  The number  $k$ is the tensorial order  where $0 \le k \le 2J$ and    $-k \; \le \; q \; \le \;k$ quantifies the coherences between the levels and the sublevels  (e.g.\ Fano 1963; Omont 1977;  Sahal-Br\'echot 1977; Landi Degl'Innocenti  \& Landolfi 2004).  
 
 In the case of interactions of singly ionized ions with neutral hydrogen, the inelastic  and super-elastic collisions between two different electronic levels  are negligible.  The indice $\alpha$  is omitted  from now on for the sake of brevity.  We apply the semi-classical theory developed by Derouich et al. (2004) and the associated numerical code to obtain, the all non zero  collisional rates of any   $p$-level.  
 
 At the solar photosphere where the SSS is formed,   the dominant collisions with neutral hydrogen are isotropic.    Due to this isotropy, the depolarization and polarization transfer rates are $q$-independent (e.g.   Sahal-Br\'echot 1977; Derouich et al. 2003a, Landi Degl'Innocenti  \& Landolfi 2004).  The  rates are obtained in the tensorial basis. We denote by $D^{k}(J)$   the  depolarization   rates  due to purely elastic collisions; the expression of $D^{k}$ is given for example by equations (7) and (9) of Derouich et al. (2003a).  $C^k(J \to J')$    are the polarization transfer rates due to  collisions between
 the initial electronic state  $(J)$ and the final state $(J')$ respectively (see Equation 3 of Derouich et al. 2003b; Landi Degl'Innocenti \& Landolfi 2004). 
 
  According to Equation (1) of Derouich et al. (2004) (see also Equation (1) of  Sahal-Br\'echot et al. (2007) and Sahal-Br\'echot  (1977)),  the variation of     $ \rho_{q}^{k} (J)$     due to isotropic collisions is:  
\begin{eqnarray} \label{eq_1}
\big[\frac{d \; \rho_q^{k} (\ J)}{dt}\big]_{coll} & = &  
  - \big[\sum_{J' \ne J} \zeta (J \to  J') + D^k(J) \big] \times \rho_q^{k} (J)  \\
&& + \sum_{J' \ne J} 
C^k(J' \to  J)   \times \rho_q^{k} (J')  \nonumber 
\end{eqnarray}
where  $\zeta (J \to  J')$ are the fine structure transfer rates  given by (Equation 5 of Derouich et al. 2003b):
\begin{eqnarray} \label{eq_1}
\zeta (J \to  J')  & = &   \sqrt{ \frac{2J'+1} {2J+1}} \times
C^0(J \to  J').    
\end{eqnarray}

 We notice that this rate equation is slightly different from the rate equation (7.101) of Landi Degl'Innocenti  \& Landolfi (2004) (page 343).   To avoid confusion, we will demonstrate that these equations are not in contradiction, they are similar.  

 In fact, as it was mentioned for example in  Derouich (2008) (see also Derouich \& Ben Abdallah (2009)),  there is a multiplicity factor equal to $\sqrt{\frac{2J'+1}{2J+1}}$ between the collisonal rates  $C^{k}  (J \to  J')$   calculated by our method (and defined for instance in  Sahal-Br\'echot  (1977);  Derouich et al. (2004); Sahal-Br\'echot et al. (2007)) and  the collisional rates  $[C^k(J \to  J')]_{LL}$ defined by  Landi Degl'Innocenti  \& Landolfi  (2004). One has:
 \begin{eqnarray} \label{eq_1}
C^{k}  (J' \to  J)  & = &   \sqrt{\frac{2J'+1}{2J+1}}   [C^k(J' \to  J)]_{LL} 
\end{eqnarray}
 \begin{eqnarray} \label{eq_1}
\sqrt{\frac{2J'+1}{2J+1}} C^{0}  (J \to  J')  & = &     [C^0(J \to  J')]_{LL} 
\end{eqnarray}
Since for the depolarization rate   one has $J=J'$ and thus $\sqrt{\frac{2J'+1}{2J+1}}=1$, the $D^k(J)$ is the same in both definitions.
According to Landi Degl'Innocenti  \& Landolfi  (2004), the  rate equation    is:
\begin{eqnarray} \label{eq_1}
\big[\frac{d \; \rho_q^{k} (\ J)}{dt}\big]_{coll} & = &  
  - \big[\sum_{J' \ne J} [C^0 (J \to  J')]_{LL} + D^k(J) \big] \times \rho_q^{k} (J)  \\
&& + \sum_{J' \ne J}  \sqrt{\frac{2J'+1}{2J+1}}
[C^k(J' \to  J)]_{LL}   \times \rho_q^{k} (J')  \nonumber 
\end{eqnarray}
It is easy to see that, using our collisional rates, the  rate equation of Landi Degl'Innocenti  \& Landolfi  (2004) (Equation 7 of this paper)  becomes exatly the same as the rate equation given in the present paper (Equation 3).

In our present paper,   to ensure continuity, we adopt  the definitions related to Equation 3 since they are known from the older work of Sahal-Br\'echot   (1977) and   they
follow  the definitions adopted in our previous papers.  In any case, this  is nothing more than a difference in deÞnitions. Nevertheless, this difference should be taken into account when writing the variation of the density matrix    in order to calculate the polarization signals correctly.  For this reason, Equation 3 is necessary for   a reader  who wants to exploit our work  paper.

In our calculations, we  consider  a simple ion with   one   optical electron  in the external shell  
with an 
orbital angular momentum $\vec{l}$ and spin $\vec{s}$ (s=1/2). The cases of complex atoms and atoms with  hyperfine structure can be easily derived as 
it is explained in the last sections
 of this paper. 

 Our results are concentrated on 
  the  $p$-states ($l$=1) case  which is frequently encountered in the SSS. In the   LS coupling scheme, the total angular momentum  $\vec{J}$=$\vec{l}$+$\vec{s}$ and one obtains $J$=1/2 or $J$=3/2. These $J$-values are associated to the states   $^2P_{J=\frac{1}{2}}$ and   $^2P_{J=\frac{3}{2}}$.

 We notice that all the collisional rates are given in s$^{-1}$. Since the impact approximation  is well satisfied, the rates are proportional to $n_{\textrm {\scriptsize H}}$ which is the neutral hydrogen density in cm$^{-3}$. They depend on the temperature $T$ given in Kelvins.   The collisional  
rates are usually  expressed as (e.g. Derouich et al. 2003a;  Derouich et al. 2004)
\begin{eqnarray} \label{eq_1}
D^k(J,T) & = &  D^k(J,T=5000K) \times    (\frac{T} {5000})^{ \frac{1-\lambda} {2}} \\
C^k(J \to J',T) & = &  C^k(J \to J',T=5000K) \times    (\frac{T} {5000})^{ \frac{1-\lambda} {2}} 
\end{eqnarray}
where $T=5000$K is the reference temperature for which collisional rates are tabulated and $\lambda$ is the so-called velocity exponent. DSB and 
 ABO found  that  $\lambda$  has a limited range of variation around 0.25 and, as a result, that collisional rates have typical   temperature dependence of $T^{0.38}$. 
 
 \subsection{Numerical results}
  Grids of collisional rates are computed at $T=5000K$ for  the effective principal quantum 
number $n^*$ of the    $p$-states in the interval [1.5, 3].  In addition, we calculated for each   $n^*$,  grids of collisional rates for Uns\"old energy $E_p$  ranged
 in the interval [-1.2, -0.1] (in atomic units).  We adopt a step size of 0.1 when varying  $n^*$ and $E_p$. As a results of our calculations, we obtain   three dimensional tables giving the   collisional rates  with $n^*$ and $E_p$.   
 
 Using these tables, we applied our Genetic Programming (GP)-based method in order to infer analytical relationships between  collisional rates, 
 $n^*$ and $E_p$. This method has been used in Derouich et al. (2015).  We minimize, in a least squares sense, the difference
between the tabulated rates  and rates obtained from the  GP  relationships.  The predicted relationships  from the GP-based model are compared with the available  data of the $p$-state of the Ca II  in order to estimate the accuracy of our results. 
  
 The collisional depolarization/transfer of polarization rates are in   the left-hand side of  the  relationships.  The second term in the right hand side   of the  relationships is a function of   $n^*$ and $E_p$.  For $J=\frac{1}{2}$, only the value $k=1$ is possible. However for $J=\frac{3}{2}$,   $0 \le k \le 3$. In addition, by definition,   $D^{k=0}(J)$=0. Thus the non-zero depolarizations rates associated to the levels  $^2P_{J=\frac{1}{2}}$ and $^2P_{j=\frac{3}{2}}$ are:
 \begin{itemize}
\item
\begin{eqnarray}  \label{eq_1}
D^{1}(\frac{1}{2}) (T=5000K)/(n_{\textrm {\scriptsize H}}  \times 10^{-9})&=&  X-\frac{Y^2}{49.} +\frac{(X-1.)}{(28.  Y-Y \times X)} \nonumber \\  
&-&  \frac{Y}{(2.-\frac{Y}{(X \times Y-2.  X)})}  
\end{eqnarray}
 where  $X=n^{*}_{p}$    and $Y=  -E_p >0$. For instance, let us consider the case of the $p$-state of the Ca II where $n^{*}_{p}=2.49 $ and $E_{p}$=-0.544.   According to the Equation 12 of Derouich et al. (2004),  $D^{1}(\frac{1}{2}) (T=5000K)$/$(n_{\textrm {\scriptsize H}}  \times 10^{-9})$=2.48  s$^{-1}$. Equation (\ref{eq_1}) gives $D^{1}(\frac{1}{2}) (T=5000K)$/$(n_{\textrm {\scriptsize H}}  \times 10^{-9})$=2.34   s$^{-1}$. Thus, the    relative error is less than 6 \% which demonstrates how good is the  precision of our GP method of fitting.  
\item
 \begin{eqnarray}  \label{eq_11}
D^{1}(\frac{3}{2}) (T=5000K) /(n_{\textrm {\scriptsize H}}  \times 10^{-9})  &=&  [5.+\frac{X}{(10. Y+\frac{5.}{2.}  Y \times X)}] \times \nonumber \\ 
\times [\frac{(3.  X+X^2)}{7.} 
+  \frac{7.}{(10.+2  X)}]  &\times&  [\frac{1.}{(Y+5.-\frac{5.}{X^2})}]
\end{eqnarray}
In the case of the $p$-state of the Ca II, the relative error on the determination of $D^{1}(\frac{3}{2})$ is  less than 4 \% with respect to the reference value given in Equation 13 of Derouich et al. (2004).
\item
 \begin{eqnarray}  \label{eq_11}
D^{2}(\frac{3}{2}) (T=5000K)  /(n_{\textrm {\scriptsize H}}  \times 10^{-9}) &=&   \frac{X^2 -7.}{Y-3 X+\frac{68.}{7.}}+X^2-Y \times X  \nonumber \\  & & + Y - \frac{Y}{(14.+7. \times Y)}  \nonumber  \\
\end{eqnarray}
With   respect to the reference value given in Equation 13 of Derouich et al. (2004), the  relative error on the determination of $D^{2}(\frac{3}{2})$  is $\simeq$ 2 \% in the case of the $p$-state of the Ca II. 
\item
 \begin{eqnarray}  \label{eq_11}
D^{3}(\frac{3}{2}) (T=5000K) /(n_{\textrm {\scriptsize H}}  \times 10^{-9})  &=&   \frac{(X^2+\frac{8.}{5.})}{(7.-\frac{X}{5.}-X+Y)}    \\  &  \times& (1.+X-\frac{Y}{(Y+2.)}-0.5)  \nonumber   
\end{eqnarray}
By comparing to the reference value of $D^{3}(\frac{3}{2}) (T=5000K)$ given in Equation 13 of Derouich et al. (2004), the  relative error  is $\simeq$ 2 \% in the case of the $p$-state of the Ca II.
\end{itemize}
Concerning the non-zero polarization transfer rates  between the levels   $^2P_{J=\frac{1}{2}}$ and $^2P_{J=\frac{3}{2}}$, only the rates $C^{0}(\frac{1}{2} \to \frac{3}{2}) $, $C^{0}(\frac{3}{2} \to \frac{1}{2}) $,  $C^{1}(\frac{1}{2} \to \frac{3}{2})$, and  $C^{1}(\frac{3}{2} \to \frac{1}{2})$ are   non-zero,
\begin{itemize}
\item
 \begin{eqnarray}  \label{eq_11}
C^{0}(\frac{1}{2} \to \frac{3}{2})  (T=5000K) /(n_{\textrm {\scriptsize H}}  \times 10^{-9})  &=& \frac{X^2}{(6.+\frac{X}{2.}-\frac{X^2}{2.}+Y)} \nonumber \\ &\times &  [X+\frac{(5.-Y)}{(5.+Y)}]  
\end{eqnarray}
By comparing to the reference value of $C^{0}(\frac{1}{2} \to \frac{3}{2})  (T=5000K))$ given in Equation 17 of Derouich et al. (2004), we found a relative error $\simeq$ 8 \% for the case of the of the $p$-state of the Ca II. Note that in Derouich et al. (2004), the polarization transfer rates are denoted by $D^{k}(J' \to J,T)$ instead of the notation  $C^k(J' \to J,T)$ adopted here.
\item
 \begin{eqnarray}  \label{eq_11}
C^{1}(\frac{1}{2} \to \frac{3}{2})  (T=5000K)  /(n_{\textrm {\scriptsize H}}  \times 10^{-9}) &=& \\  \frac{\frac{(X+1.)}{(X-1.) \times (X+Y)}+\frac{(2.+Y)}{(5.-X)}+6.}{(X^2-Y \times X-23./2.)}  \nonumber 
\end{eqnarray}
By comparing to the reference value of $C^{1}(\frac{1}{2} \to \frac{3}{2})  (T=5000K))$ given in Equation 17 of Derouich et al. (2004), we found a relative error is less than 3 \%.  
\end{itemize}
Only the excitation  collisional transfer rates $C^{0}(\frac{1}{2} \to \frac{3}{2})$ and $C^{1}(\frac{1}{2} \to \frac{3}{2}) $ are given. However, it is 
straightforward to retrieve the values of the   deexcitation  collisional rates 
 $C^{0}(\frac{3}{2} \to \frac{1}{2}) $  and  $C^{1}(\frac{3}{2} \to \frac{1}{2}) $ by applying the detailed balance relation:
 \begin{eqnarray}  \label{eq_11}
C^k(J_u \to J_l, T) &=&     \frac{2J_l+1}{2J_u+1}  \exp \left(\frac{E_{J_u}-E_{J_l}}{k_BT}\right) \; \; C^k_I(J_l \to J_u, T)
\end{eqnarray}
where $J_l$=1/2 (lower  level) and     $J_u$=3/2   (upper  level); $E_{J}$ being the energy of the level ($J $) and $k_B$ the Boltzmann constant.

Thanks to the relationships given here, any   collisional rates can be obtained by simply determining the value of $n^*$ and $E_p$, i.e. by performing the step 1 as explained in the previous section without the need to step 2. This is the main result of this work.

\section{Hyperfine structure}
In the typical solar conditions where the temperature is about $5700$K,  the   inverse of the typical time duration of a collision  between a hydrogen atom   and the perturbed   ion  is $1/\tau$ $\sim$ 10$^{13}$ s$^{-1}$.  In these conditions,  the hyperfine  splitting is usually much smaller than $1/\tau$ and therefore one can assume that the nuclear spin is conserved during the collision\footnote{It is important however to  not confuse this condition with the fact that the  statistical equilibrium    must be
solved for the hyperfine levels when the inverse of the lifetime
of the level is smaller than the hyperfine splitting, i.e. the
hyperfine levels are separated.}. This is the     frozen nuclear spin   approximation.
   
In the framework of the    frozen nuclear spin approximation, the depolarization and polarization transfer rates of hyperfine  levels are given as a linear combination of the  rates $D^k(J)$ and $C^k(J \to  J') $  associated to the    levels of the fine structure  (e.g. Nienhuis 1976 and Omont 1977). 
Using the results of this work and after calculating simple coefficients of the  linear combination which are illustrated in Derouich et al. (2005b) and given 
by Equations (4.6) 
of Nienhuis (1976) and (4.48) of Omont (1977), it is possible   to perform sufficiently accurate computation of the needed collisional depolarizing and polarization transfer  rates   for all levels of    ions with hyperfine structure. 

\section{Complex atoms}
Our relationships given by Equations 7--12 can be directly applied to obtain depolarization and polarization transfer rates
of the lines of singly   ionized ions like Be II, Mg II, Ca II, Sr II, and 
Ba II in their $p$-states since they are simple ions because they have 
only one valence electron above a filled subshell. In contrast, 
the electronic conÞguration of a complex ion has one valence 
electron above an incomplete (open) subshell within the core. 
 Derouich et al. (2005a) provided, from a conceptual and numerical point views, the physical 
model  allowing, for the first time, the determination of  the depolarization and polarization transfer rates of 
complex ions and atoms by using the rates of simple atoms. More details   are also given in Derouich et al. (2005b), Derouich \& barklem (2007) and Sahal-Br\'echot et al. (2007).
In these works, we demonstrated  that the  depolarization and polarization transfer rates of complex atoms can be 
written as a linear combination of the rates of simple atoms. Therefore, it is straightforward to use our Equations  7--12   to infer rates of complex atoms 
by simply calculating some algerba coefficents whose expressions are given in Derouich et al. (2005b), Derouich \& barklem (2007) and Sahal-Br\'echot et al. (2007).    
\section{Conclusion}
Derouich et al. (2004) proposed a semi-classical theory allowing the determination of the  depolarization and polarization transfer rates of any simple ion, but 
 one must apply the theory and the collisional numerical code in order to calculate these rates for each level of each ion. This makes the use of the results of 
   Derouich et al. (2004) by the solar physics community  rather limited.
 The purpose of the present paper is   to generalize the semi-classical theory of Derouich et al. (2004) by giving general relationships   allowing the determination of the  depolarization and polarization transfer rates of any simple ion  without the need to perform collisional calculations for each level. In addition, these relationships  can be used to determine the depolarization/polarization transfer rates of the levels of hyperfine structered ions. Furthermore,  these relationships allow sufficiently accurate determination of the rates associated to complex ions whose collisional rates cannot be presently obtained via   standard quantum chemistry methods. We notice that the spectral lines of such  ions   show significant polarization peaks in many spectral lines of the SSS (see the atlases by Gandorfer (2000, 2002, 2005)). 
 
 All the results of  this paper are performed thanks to powerful  GP techniques. Genetic algorithms have been found to be
very good at determining the global minimum in a space with many local minima.
  The details of the GP approach can be found in Derouich et al. (2015) and references therein.

   \section*{Acknowledgements}
I thank  Dr. P. Barklem for stimulating scientific discussions and ongoing collaboration. I would like to thank referee 1  for the detailed constructive comments.

\end{document}